\documentclass{article}

\usepackage{epsfig,amsfonts}

\begin{document}

\newcommand{\comentario}{\scriptsize}

\newcommand{\bbox}[1]{\mbox{\boldmath $#1$}} 
\newcommand{\sbbox}[1]{\mbox{\scriptsize\boldmath $#1$}}
\newcommand{\RP}[1]{\mathbf{R}\mathrm{P}^#1}
\newcommand{\SO}[1]{\mathrm{SO}(#1)}
\newcommand{\Orth}[1]{\mathrm{O}(#1)}
\newcommand{\Tc}{T_\mathrm{c}}
\newcommand{\betac}{\beta_\mathrm{c}}
\newcommand{\betas}{\beta_\mathrm{s}}
\newcommand{\gammas}{\gamma_\mathrm{s}}
\newcommand{\etas}{\eta_\mathrm{s}}
\newcommand{\chis}{\chi_\mathrm{s}}
\newcommand{\xis}{\xi_\mathrm{s}}
\newcommand{\nus}{\nu_\mathrm{s}}
\newcommand{\lambdas}{\lambda_\mathrm{s}}
\newcommand{\kappas}{\eta_\mathrm{s}}
\newcommand{\stag}{\mathrm{s}}
\newcommand{\tr}{\mathop{\mathrm{tr}}}
\newcommand{\e}{\mathrm{e}}
\renewcommand{\d}{\mathrm{d}}

\title{Critical properties of the Antiferromagnetic \\ 
$\RP2$ model in three dimensions}

\author{H.~G.~Ballesteros, L.~A.~Fern\'andez,\\
        V.~Mart\'{\i}n-Mayor, and A.~Mu\~noz Sudupe.\\
\\
\it     Departamento de F\'{\i}sica Te\'orica I, 
        Facultad de CC. F\'{\i}sicas,\\ 
\it     Universidad Complutense de Madrid, 28040 Madrid, Spain.\\
{\small e-mail: {\tt hector,laf,victor,sudupe@lattice.fis.ucm.es}}}

\maketitle

\begin{abstract}
We study the behavior of the antiferromagnetic $\RP2$ model in
\hbox{$d=3$}. The vacuum structure is analyzed in the critical and low
temperature regions, paying special attention to the spontaneous
symmetry breaking pattern. Near the critical point we observe a full
breakdown of the $\Orth3$ symmetry of the action.  Several methods for
computing critical exponents are compared. We conclude that the most
solid determination is obtained using a measure of the correlation
length. Corrections-to-scaling are parameterized, yielding a very
accurate determination of the critical coupling and a 5\% error
measure of the related exponent. This is used to
estimate the systematic errors due to finite-size effects.
\end{abstract}

\bigskip

Nonlinear sigma model.
Antiferromagnetism.
Critical exponents.
Finite size scaling.


\newpage

\section{Introduction.}

Antiferromagnetic, (AF) interactions give rise to very interesting
properties in statistical systems. Specifically, the symmetries of the
broken AF phase can be very different from their ferromagnetic
counterparts. Ground states with frustration or
disorder are common in these systems.

oThe interest of AF models extends from the study of some physical
problems where the antiferromagnetism appears naturally to the more
theoretical aspects of new Spontaneous Symmetry Breaking (SSB)
patterns, or Universality Classes.
Moreover, attention has been recently paid to AF models in four
dimensions~\cite{POLONYI,ISING4AF}.  There, one could even hope to gain
some insight on the puzzling problem of the non perturbative
formulation of interacting field theories.

In this paper we will be concerned with the $\RP2$ model in three
dimensions.  Its ferromagnetic sector has been extensively used to
study liquid crystals~\cite{LIQUIDCRYS}. The AF one has been also
studied~\cite{SHROCK,ROMANO,LETTER} as it presents an unfrustrated
ground state with nonzero disorder. It seems to belong to a class of
models with a SSB pattern of type
\mbox{$\SO3\times\SO2/\SO2$}~\cite{LETTER}. In condensed matter physics
it is not rare to find systems with an equivalent SSB pattern, like
super-fluid $^3$He~\cite{HETRES}, helical~\cite{DIEP} and
canted~\cite{KAWAMURA} spin systems and some frustrated quantum AF
Heisenberg models~\cite{DOMBRE}, the latter being specially
interesting because of their possible relation with High Temperature
Superconductivity~\cite{HTSC}.

A radical reading of the Universality Hypothesis, suggests that the
critical properties of a model are completely fixed by the space
dimension and by its low and high temperature phase symmetry groups,
$H$ and $G$ respectively.  Even more, it is also generally assumed
that models with locally isomorphic manifolds $G/H$ have identical
critical behaviors, regardless of the global properties.  This
scenario has been theoretically challenged in ref.~\cite{ZUMBACH1},
where it has been suggested that the massive modes, not fixed by the
local properties of $G/H$, could change the numerical values of the
critical exponents. 

The above SSB pattern has been studied in
perturbation theory~\cite{AZARIA}, where the main conclusions reached
are that the only possible critical points in three dimensions are
second-order, either with the Mean-Field (MF) exponents or
with those of the $\Orth4$ sigma model, or first-order.

AF interactions frequently produce first-order
transitions~\cite{O3A,ISING4AF}, although with a large correlation
length (weak-first-order transitions) that are characterized by the
(apparent) critical exponents: $\nu=0.5$, $\gamma/\nu=2$,
$\alpha/\nu=1$~\cite{WFO}. These exponents are found in
ref.~\cite{ZUMBACH2} for the Stiefel manifold $V_{3,2}$.  However,
other Monte Carlo (MC) simulations for helical and canted spin
models~\cite{DIEP,KAWAMURA,STA}, have yielded exponents not far but
incompatible with those of a weak-first-order transition.

On the other hand, the $\RP2$ antiferromagnetic model has a critical
point with clearly different from weak-first-order exponents
($\nu\approx 0.78$, $\alpha/\nu\approx -0.44$)~\cite{LETTER} and 
very close to those reported for $\Orth4$~\cite{O43D}, although
hardly compatible with them.

The actual relevance of the model in experimental situations is
difficult to know. It is an uneasy task to compare the critical
exponents of this model with experimental data. In fact, the results
for spin systems are not very clear, even regarding the order of the
phase transitions encountered (see ref.~\cite{DIEP} for a review). For
the super-fluid $^3$He, the possibility of facing our critical
exponents with experimental measures is remote, the critical region
being so small that it seems experimentally
inaccessible~\cite{WOELFLE}.

In this paper we present a more detailed exposition of the studies of
ref.~\cite{LETTER}, completed with new analysis of the MC data.  Our
statistics have also increased by an amount of about a 50\%.  The high
statistics has allowed a thorough study of the autocorrelation times,
both exponential and integrated, for several observables.  We are 
able to parameterize the corrections-to-scaling.  As a consequence, we
get a high precision determination of the critical point, and we
measure the first correction-to-scaling exponent with an error of
5\%. We use this exponent to extrapolate the critical exponents to their
infinite-volume values, and get an estimation of the systematic
errors. In addition, we carry out a study of the ordered phase both
numerically and analytically. We find a low temperature phase with a
global $\Orth2$ symmetry and, from finite-size scaling analysis,
evidence of a breaking of this symmetry close to the transition.  The
MF approximation is considered at the critical point
which already shows the extreme complexity of the symmetry-breaking
pattern in that region.  We also analyze the continuum limit of the
model obtaining the $\SO3\times\SO2/\SO2$ non linear sigma model.

We start in section \ref{MODEL} by defining the model and the
observables, and discuss some theoretical aspects and approximations.
The MC algorithms used and the dynamics of the simulation are
described in section \ref{MCSIM}. In section \ref{EXPONENTS} we
present a determination of the critical exponents using
finite-size scaling techniques.  The reliability of a parameterization
of the corrections-to-scaling is considered in section \ref{CORRSCAL}.
Finally sections \ref{VACUUM} and \ref{LOWTEMP} are devoted to discuss
the vacuum structure of the broken phase. The conclusions are in
section \ref{CONCLUS}.

\section{The Model.}\label{MODEL}

We will consider a system of spins $\{\bbox{v}_i\}$ taking values in
the sphere $\mathrm{S}^2 \subset \mathbf{R}^3$ and placed in the nodes
of a cubic lattice. The interaction is defined by the action
\begin{equation}
{\mathcal S}=-\beta \sum_{<i,j>} (\bbox{v}_i\cdot\bbox{v}_j)^2 ,
\label{HAMILTONIAN}
\end{equation}
where the sum is extended over all pairs of nearest neighbor sites.
The partition function is constructed as
\begin{equation}
{\mathcal Z}=\int \left(\prod_i \d\bbox{v}_i\right) \e^{-\mathcal S},
\end{equation}
where the usual $\Orth3$-invariant measure is used.

The action and the measure are invariant under a local $\mathbf{Z}_2$
transformation ($\bbox{v}_i\to-\bbox{v}_i$). As this is a local
symmetry, Elitzur's Theorem 
guarantees its
preservation even after a SSB.  This means that the natural variable
for the model is in an equivalent class of the quotient group
$\RP2=\mathrm{S}^2/\mathbf{Z}_2$, the two-dimensional real projective
space. The action (\ref{HAMILTONIAN}) can be rewritten in terms of the
$\RP2$ variables, which in turn can be expressed as the tensorial
product of the original spins
$\bbox{\tau}_i=\bbox{v}_i\otimes\bbox{v}_i$,
\begin{equation}
{\mathcal S}=-\beta \sum_{<i,j>} \tr \bbox{\tau}_i \bbox{\tau}_j.
\end{equation}
As the measure for the $\bbox{\tau}$ is much more complex than for the
vectors, it is generally more convenient to work directly with the
latter.

\subsection{Observables and measures.}

The observables to be measure in a Monte Carlo simulation on a finite
lattice must be scalars under the $\Orth3$ global symmetry group
and obviously also under the local $\mathbf{Z}_2$ group.  Let us
construct the traceless tensorial field $\mathbf T_i$, whose
components are

\begin{equation}
\mathrm T_i^{\alpha\beta}=
 v_i^\alpha v_i^\beta - \frac{1}{3}\delta^{\alpha\beta} .
\label{TENSORFIELD}
\end{equation}

The Fourier transform in a $L\times L \times L$ lattice can be defined
as follows:
\begin{equation}
\widehat{\mathbf T}_{\sbbox p}=\sum_{\sbbox r \in L^3} 
\e^{-i \sbbox p\cdot \sbbox r}\ {\mathbf T}_{\sbbox r} ,
\end{equation}
where we denote by $\bbox r$ the position vector of the site, and by
$\bbox p$ the momentum ($\bbox p=(2 \pi n_x/L,2 \pi n_y/L, 2 \pi
n_z/L)$, with $n_i=0,\ldots,L-1$).

In principle, we are only interested in the magnetizations associated
with a fixed momentum. In addition to the (normalized) zero momentum
magnetization
\begin{equation}
{\mathbf M}=\frac{1}{V}\sum_{x,y,z} {\mathbf T}_{(x,y,z)}
  =\frac{1}{V}\widehat{\bf T}_{(0,0,0)},
\label{MAG}
\end{equation}
it is interesting to study the staggered magnetization related with
states of period 2 under translations,
\begin{equation}
{\bf M}_\mathrm{s}=\frac{1}{V}\sum_{x,y,z} (-1)^{x+y+z} 
{\mathbf T}_{(x,y,z)} =
\frac{1}{V}\widehat{\mathbf T}_{(\pi,\pi,\pi)},
\label{MAGS}
\end{equation}
that is the (normalized) difference between the magnetization of the
even sites, defined as those where $x+y+z$ is even, minus that of the
odd ones, defined analogously.

Other finite momenta (nonzero in the $L\to\infty$ limit) could be
necessary if the vacuum would present higher period symmetries.
However we have not found any reason, neither analytical nor numerical, to
expect this.
 
We compute the magnetization and the susceptibility respectively as
\begin{equation}
M=\left\langle \sqrt{\tr {\mathbf M}^2}\right\rangle,
\end{equation}
\begin{equation}
\chi=V \left\langle \tr {\mathbf M}^2\right\rangle .
\end{equation}
It is also useful to consider the quantity
\begin{equation}
\kappa=\frac{\langle \left(\tr \mathbf{M}^2\right)^2\rangle}
              {\langle \tr \mathbf{M}^2\rangle^2},
\label{PSEUDOBINDER}
\end{equation}
which is directly related with the Binder parameter~\cite{BINDER}.

We use the \emph{second momentum} correlation length because it is
easier to measure than the exponential (physical) one, but it is
expected to have the same scaling behavior at the critical
point~\cite{CORREL,KIM},

\begin{equation}
\xi=\left(\frac{\chi/F-1}{4\sin^2(\pi/L)}\right)^{1/2},
\label{XI}
\end{equation}
where $F$ is defined, using the shorthand notation $|\widehat{\mathbf
T}|^2=\widehat{\mathbf T}\widehat{\mathbf T}^\dagger$, as
\begin{equation}
F=\frac{1}{3V} 
\left\langle \tr\left(
 \left|\widehat{\mathbf T}_{(\frac{2\pi}{L},0,0)}\right|^2
+\left|\widehat{\mathbf T}_{(0,\frac{2\pi}{L},0)}\right|^2
+\left|\widehat{\mathbf T}_{(0,0,\frac{2\pi}{L})}\right|^2
\right)\right\rangle.
\end{equation}

Analogously we define the associated quantities for the staggered 
observables. 

We also measure the nearest neighbor energy
\begin{equation}
E_1= \frac {1}{3V} \sum_{<i,j>} (\bbox{v}_i\cdot\bbox{v}_j)^2 ,
\label{E1}
\end{equation}
and the next-to-nearest neighbor energy 
\begin{equation}
E_2= \frac {1}{6V} \sum_{\ll i,j\gg} (\bbox{v}_i\cdot\bbox{v}_j)^2 ,
\label{E2}
\end{equation}
which is useful to study the $\Orth3$ broken phase.
 
The measures of $E_1$ have been also used to calculate the $\beta$
derivatives through its connected correlation with every observable.

\subsection{The ground state.}

When $\beta\gg 0$ the ground state corresponds to a configuration in
which all the spins are aligned in a given direction. This state is
thus $\Orth2$ invariant. The thermal fluctuations do not destroy this
ordering until $\beta\sim 1.3$, where a first-order phase transition
occurs.  This system has been used for modeling
liquid crystals~\cite{LIQUIDCRYS}.

For negative $\beta$, the ground state corresponds to a highly
degenerated non-frustrated state where all spins are orthogonal to
their neighbors.

The zero energy configurations that dominate the measure are those in
which the even (odd) sublattice is aligned, and the spins in the odd
(even) one lie randomly in the orthogonal plane.  It is easy to check
that other subsets, such as, for example, those with spatial
periodicity 3~\cite{SHROCK} or without any periodicity, have a
relatively null contribution to the measure.

When $-\beta$ is large but finite, the fluctuations effects
must be taken into account.  Let us consider a spin of the aligned
sublattice surrounded by its neighboring spins lying in the orthogonal
plane. Should they prefer to orientate in a given direction of the
plane, that tendency would permit a larger fluctuation of the spin of
the aligned lattice.  As a consequence, there is an induced
interaction between second-neighbor spins in the plane-aligned
sublattice.  One can think of a $\RP1$ model (that is equivalent to an
$\Orth2$ model) in a face centered cubic lattice with a nonzero
coupling.  The important question is if the induced coupling is large
enough to produce an $\Orth2$ phase transition.  The Monte Carlo
results show that, close to the antiferromagnetic phase transition,
the $\Orth2$ symmetry is also broken, but it is restored for lower
values of $\beta$. We address to sections \ref{VACUUM} and
\ref{LOWTEMP} for a detailed discussion on this subject.

\subsection{The Mean-Field Approximation.}

A standard tool in statistical mechanics is to study the system, close
to the critical point, by means of the MF approximation. This
analysis is known to improve as the system dimensionality grows. It is
also useful to know the predictions of a MF calculation since
any deviation may be interpreted as a sign of non-triviality.

We shall use the formulation of the MF approximation
as a variational principle (see for instance \cite{LEBELLAC}).  It can
be stated as follows: let us consider a configuration space, and the
Boltzmann measure $\mathrm D\mu = ({\mathcal Z}(T))^{-1}
\exp({\mathcal H}(\mu)/T) d\mu$, where $T$ is the temperature.  For
any other Hamiltonian ${\mathcal H}_\rho(\mu)$, the following
inequality holds,
\begin{equation}
\mathcal{F} \leq \mathcal{F}_{\rho} + \langle{\mathcal H}-{\mathcal
H}_{\rho}\rangle_{\rho}= {\mathcal G}_{\rho},
\label{DES}
\end{equation}
where by $\mathcal{F}$ we mean $-T\log {\mathcal Z}(T)$, by $\langle
O\rangle_{\rho}$ we refer to the mean value of the operator $O$,
calculated with the Boltzmann measure of the Hamiltonian
\mbox{${\mathcal H}_\rho(\mu)$, and ${\mathcal F}_\rho$} is the free
energy correspondent to ${\mathcal H}_\rho(\mu)$.

To study the action (\ref{HAMILTONIAN}) we have used the Hamiltonian
\begin{equation}
{\mathcal H}_\rho= 
-\sum_{i\ \mathrm{even}}\tr\bbox{\tau}_i \mathbf Q_{\mathrm e} 
-\sum_{i\ \mathrm{odd}} \tr\bbox{\tau}_i \mathbf Q_{\mathrm o}
\end{equation}
where $\mathbf Q_\mathrm{e,o}$ are tensorial mean fields.  The
usual procedure consists on choosing the parameters of those fields
in order to minimize ${\mathcal G}_{\rho}$.  For the sake of
simplicity, we restrict ourselves to the case of commuting $\mathbf
Q_\mathrm{e,o}$.  It is easy to show that in this way ${\mathcal
G}_{\rho}$ depends on four parameters in addition to the coupling.
The mean field equations are very complex, but they can be simplified
in the neighborhood of a continuous transition.

We report here the qualitative results addressing the reader to
appendix \ref{MFAPPEN} for details. Regarding the critical exponents,
we find the expected results $\alpha=0$ and $\betas=1/2$, where the
latter is the exponent related to the staggered
magnetization. However, the exponent associated with the
non-staggered magnetization turns out to be $\beta=1$.  Other
interesting result regards the structure of the ordered vacuum.
Performing a Ginzburg-Landau expansion of $\mathcal{G}_\rho$ on
$M_\stag$ we observe that, up to fourth order, an $\Orth{2}$ symmetric
vacuum has the same free energy as a broken one.  The sixth order term
favors the unbroken vacuum.  However, this quasi degeneracy suggests
that the problem of the $\Orth{2}$ breakdown is beyond the reach of
the MF approximation.

\subsection{The continuum limit of the $\Orth2$ broken model.}

In reference~\cite{AZARIA} the continuum limit of systems with a SSB
pattern of type $\SO3\times\SO2/\SO2$ was considered with the
conclusions regarding the order of the transition mentioned in the
introduction.

We have found that as $\beta\to -\infty$ (see section~\ref{LOWTEMP}
for details) the vacuum is $\Orth{2}$ symmetric, and it cannot be
described with the above mentioned SSB pattern.  However, in the
critical region, the remaining $\Orth{2}$ symmetry seems to be broken,
and then, we should consider a pattern of type $\Orth3/\{1\}$, which
is locally isomorphic to $\SO3\times\SO2/\SO2$.  Although
in the critical region one cannot assume infinitesimal fluctuations,
it is possible to relate our model with one in which we add an
explicit (large enough) ferromagnetic interaction between
second-neighbors. That model presents a vacuum with the same
symmetries as the one with the action (\ref{HAMILTONIAN}) in the
broken critical region.  It can be shown (see appendix B for details)
that, in the continuum limit, the action of this model can be written
as
\begin{equation}
{\mathcal S}= 
\int \d^3 {\bbox x}\sum_\mu[P(R^T\partial_\mu R)^2],
\label{CONTHAMIL}
\end{equation}
where $R({\bbox x})$ is an $\SO3$-valued field and $P$ is a diagonal
matrix of couplings of type $\{g_1,g_1,-g_2\}$.

The action (\ref{CONTHAMIL}) is in the class studied in
ref.~\cite{AZARIA}.

\subsection{The low temperature effective model.}\label{FLATMODEL}

For $-\beta$ very large, it is useful to consider a limiting model as
follows. Let us suppose that the fully aligned sublattice, say in the
$z$ axis direction, is the even one.  Writing the even and odd spins
as

\begin{eqnarray}
\qquad\qquad\quad\bbox v^\mathrm{e}&=&(v^x,v^y,\sqrt{1-(v^x)^2-(v^y)^2}),\\ 
\bbox v^\mathrm{o}&=&
(\sqrt{1-(v^z)^2}\cos\varphi,\sqrt{1-(v^z)^2}\sin \varphi,v^z),\label{VFLAT}
\end{eqnarray} 

respectively, we obtain the approximate action ($\alpha=x,y,z$)
\begin{equation}
{\mathcal S}=-\displaystyle\beta 
        \sum_{\stackrel{\ll i,j\gg}{i\ \mathrm{even}}} 
(v^x_i\cos \varphi_j +v^y_i\sin \varphi_j+v^z_j)^2+O((v^\alpha)^4).
\label{FLATHAMIL}
\end{equation}
Rescaling the spins as $\beta (v^\alpha)^2 \rightarrow (\hat
v^\alpha)^2$ we conclude that in the large $-\beta$ limit the
$\varphi$'s 
correlation functions become $\beta$-independent.

The action (\ref{FLATHAMIL}) is easy to simulate, although one should
be aware that the alignment direction in a finite lattice rotates
(Goldstone modes).  To avoid this problem, a global rotation after
every Metropolis sweep should be performed.  We are particularly
interested in the correlation of the second neighbor angles: $\langle
\cos^2(\varphi_i-\varphi_j)\rangle=E_2$.  In section \ref{LOWTEMP} we
check numerically that this model smoothly joins with $\RP{2}$ at
large $-\beta$.

\section{MC simulations.}\label{MCSIM}

We have used a standard Metropolis algorithm for the update. The
fluctuation of the spins in the critical region is very large and one
can choose as  change proposal a value almost independent of the
initial state. In fact, an uncorrelated spin is accepted with
a 30\% of probability. We have used an almost independent proposal but
with 3 hits to obtain about a 70\% of mean acceptance.

We have also checked the efficiency of cluster algorithms~\cite{SW} as
the model is suitable for applying the embedding
procedure~\cite{WOLFF}.  Unfortunately, as in most antiferromagnetic
systems, the usual cluster methods do not reduce the
exponent $z$ for the autocorrelation time (AT). We have studied both the
Swendsen-Wang's method and the Wolff's single cluster version.  The
system always presents a cluster with around a 65\% of the spins, and
the remaining 35\% forms very small clusters (only about a 1\% contain
more than 10 spins).  The results regarding the efficiency are
slightly worse than those from the Metropolis method.

The simulations have been done distributed over several computers
based on ALPHA, SPARC and Pentium processors. The total computer time
employed has been the equivalent of 18 months of ALPHA AXP3000. We
measure every 10 sweeps and store individual measures to extrapolate
in a neighborhood of the simulation coupling by using the spectral
density method~\cite{FALCIONI,FERRSWEN}.  The number of performed
sweeps in the critical region is displayed in
table~\ref{AUTOCORTAB}. Some other shorter runs have been done at the
peaks of the connected staggered susceptibility, and in the broken phase in
order to study the vacuum.

Let us first consider the statistical quality of our data.  Following
the notation of reference~\cite{SOKALCREUTZ}, we compute the
unnormalized autocorrelation function
\begin{equation}
C_{O}(t)=\frac{1}{N-t}\sum_{i=1}^{N-t} O_i O_{i+t}-\mu^2_O ,
\end{equation}
where ${O_i}$ are successive measures for the operator $O$, $\mu_O$ is
the mean value of $O$ and $N$ is the number of measures.  We define
also the normalized autocorrelation function as
\begin{equation}
\rho_{O}(t)=\frac{C_{O}(t)}{C_{O}(0)} .
\end{equation}

The statistical error in the measure of $O$ is proportional to
$\sqrt{2\tau_O^\mathrm{int}/N}$ where the integrated AT
  $\tau_O^\mathrm{int}$ can be obtained from
\begin{equation}
\label{TAU}
\widehat\tau_O^\mathrm{int}(t)=\frac{1}{2}+\sum_{t'=1}^{t}\rho_{O}(t')
\end{equation}
for large enough $t$.  In practice this value is selected
self-consistently, usually as
$t=6\widehat\tau_O^\mathrm{int}(t)$.

\begin{table}[b]
\caption{Number of Monte Carlo sweeps performed for different lattice
sizes. Measures have been taken every 10 sweeps.  The integrated
AT (in sweeps) for both magnetizations and for the
energy are also displayed.  We have discarded in each case about
$200\tau_{\chis}$ iterations for thermalization.}
\medskip
\hrule\hrule
\begin{tabular*}{\hsize}{@{\extracolsep{\fill}}rcccc}
$L$ & MC sweeps($\times 10^6$) 
& $\tau_{\chis}$ &$\tau_\chi$ &$\tau_{E_1}$\\\hline
6       & 6.71   &  7.37(3)  &5.81(2)   &6.02(2)       \\
8       & 17.07  & 11.41(4)  &7.41(3)   &7.30(3)       \\
12      & 6.51   & 24.9(2)   &13.23(12) &10.68(11)     \\
16      & 22.14  & 44.5(3)   &22.27(14) &14.41(8)      \\
24      & 8.77   &107(3)     &51.3(10)  &24.7(6)       \\
32      & 28.51  &175(6)     &90.0(8)   &34.8(3)       \\
48      & 3.93   &410(20)    &222(10)   &60(3)
\end{tabular*}
\hrule\hrule
\label{AUTOCORTAB}
\end{table}

\begin{figure}[t]
\epsfig{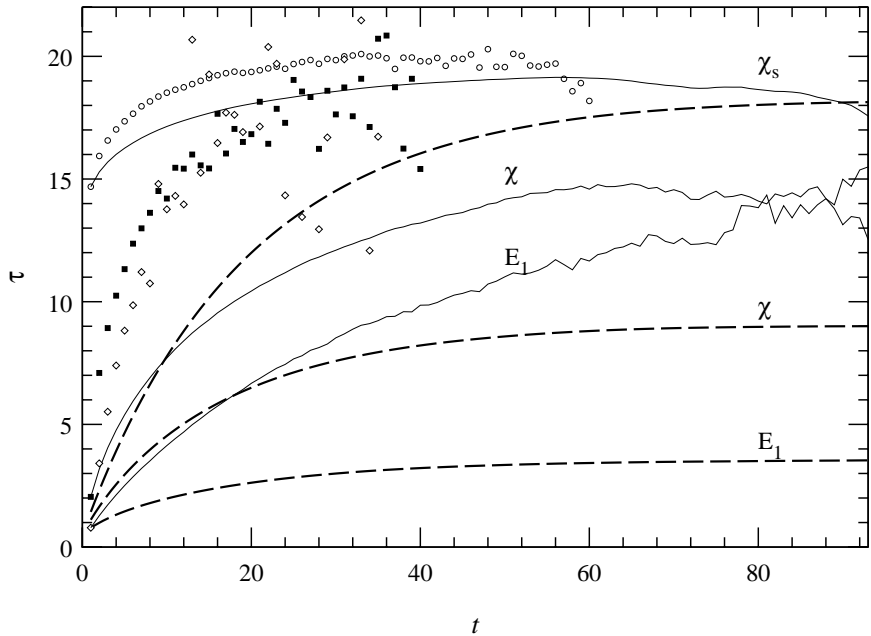}
\caption{Autocorrelation time (in measures) estimators for
$\chis$, $\chi$ and $E_1$ in a $L=32$ lattice at $\beta=-2.41$. The
dashed line correspond to the integrated time, the solid one to
$\widehat\tau_O^\mathrm{exp(1)}$, and the symbols to
$\widehat\tau_O^\mathrm{exp(2)}$ (circles, squares and diamonds for
$\chis$, $\chi$ and $E_1$, respectively).}
\label{AUTOCORFIG}
\end{figure}

In table \ref{AUTOCORTAB} we summarize the results of the simulations
at $\beta=-2.41$. In the \mbox{$L=24$} lattice, however, two thirds of
our data come from a simulation at \mbox{$\beta=-2.4$}.  We observe
that the integrated AT for the staggered susceptibility (with
$t=6\widehat\tau^\mathrm{int}_{\chis}$) grows quadratically with the
lattice size, as expected ($z$ exponent equal to 2). The AT for the
other operators in the table are smaller.  We should point out (see
fig.~\ref{AUTOCORFIG}) that we need a larger window to obtain a stable
value for those cases. Specifically we have used
$t=10\widehat\tau^\mathrm{int}_{\chi}$ and
$t=20\widehat\tau^\mathrm{int}_{E_1}$ respectively.

As the number of measures that we have performed is very large
compared with the AT
($10^5\tau_{\chis}^\mathrm{int}$ for $L=32$ and
$10^4\tau_{\chis}^\mathrm{int}$ for $L=48$), we can discard some
hundreds of $\tau_{\chis}^\mathrm{int}$ for a safe thermalization, and
make bins of consecutive data of that size to assure that they are
uncorrelated.

However, it is interesting to look at the exponential AT to know if
there is some information that remains at time scales larger than the
integrated AT.  As the exponential AT measures the
interval that the system remains out of equilibrium along a MC simulation,
only when the number of MC iterations is much greater than the
exponential AT, we can have confidence on the obtained measures. We
do not need to assume, as is usually done, that the exponential AT
is of the same order of the integrated one, as our statistics is good
enough for a reliable estimation of the latter.

The exponential AT for the operator $O$ can be measured
as the $t\to \infty$ limit of
\begin{equation}
\widehat\tau_O^{\mathrm{exp}(1)}(t)=\frac{-t}{\log \left( \rho_{O}(t)
\right)}.
\end{equation}
We should take the supreme of these quantities for all operators.

Usually it is not easy to find a clear $t\to \infty$ limit, since for
finite $t$ there is a systematic error due to the faster modes
contributions. In that case, it is useful to study the (very noisy)
estimator
\begin{equation}
\widehat\tau_O^{\mathrm{exp}(2)}(t)=\log
        \frac{\rho_{O}(t)}{\rho_{O}(t+1)}.
\end{equation}

In fig.~\ref{AUTOCORFIG} we show $\widehat\tau_{\chis}^\mathrm{int}$,
$\widehat\tau_{\chi}^\mathrm{int}$, and
$\widehat\tau_{E_1}^\mathrm{int}$ in $L=32$ in the critical region. In
all cases a plateau is clear.  The estimators
$\widehat\tau_O^{\mathrm{exp}(1,2)}$ for all the three quantities are
also plotted.  We observe that  for $\chis$ the three
estimators are almost equal. For the other two cases, there is no a
visible plateau for $\tau_O^{\mathrm{exp}(1)}$, but the estimator
$\tau_O^{\mathrm{exp}(2)}$ seems to stabilize at a value near that
corresponding to $\chis$. We remark that just with the data from $E_1$
or $\chi$ we would conclude that the AT for this
lattice size is about $200$ sweeps, that is about 6 times the
integrated AT for $E_1$.

A further check of statistical independence can be done by studying
the errors computed with bins of data of increasing size. This could
allow us to observe AT at scales much greater than the
previous.

\begin{figure}[t]
\epsfig{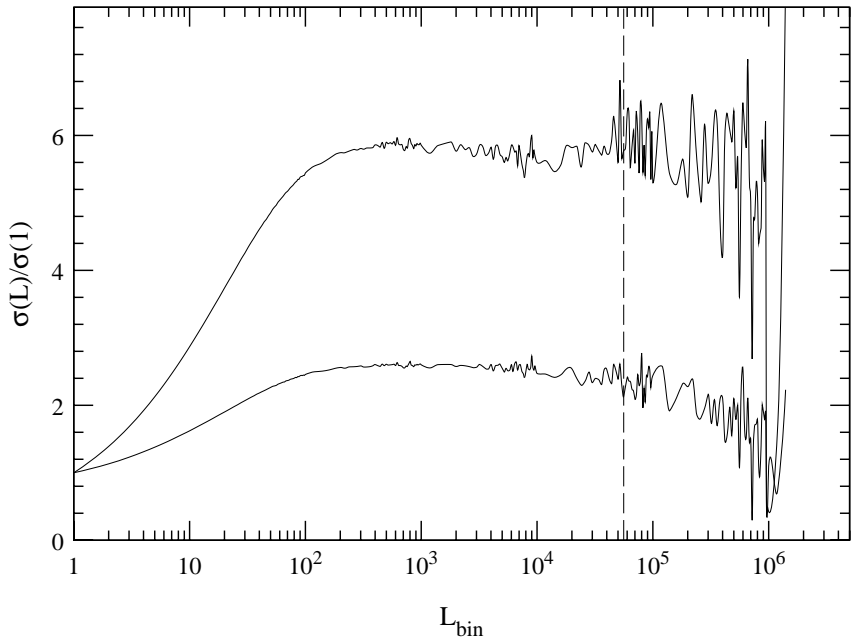}
\caption{Statistical errors of $\chis$ (upper curve) and $E_1$ (lower
curve) as a function of the bin size (in measures) for $L=32$. The
vertical dashed line corresponds to the bin size used in the
analysis.}
\label{BINS}
\end{figure}

In fig.~\ref{BINS} we show the statistical errors for the energy and
the staggered susceptibility in $L=32$ as a function of the length of
the bins, using all the statistics. When their size is not big enough,
the data in different bins are correlated and so, we see a growing
error. When the size of the bins is enough to consider them
independent, there is a plateau, that indicates the correct value for
the statistical error.  We can estimate the integrated
AT from the quotient between the value of the error at large
$L_\mathrm{bin}$ and at $L_\mathrm{bin}=1$, since
$\sigma(L_\mathrm{bin})/\sigma(1)\to\sqrt{2\tau^\mathrm{int}}$.  Using
this procedure we obtain $\tau_{\chis}^\mathrm{int}=172$
sweeps and $\tau_{E_1}^\mathrm{int}=34$ sweeps, in good
agreement with the values shown in table~\ref{AUTOCORTAB}.

From these figures we also observe that our choice of the number of
bins, 50, is completely safe, allowing an accurate estimation of the
statistical errors. We have chosen a large number of bins to ensure a
10\% precision in the statistical error determination.

From our results we find very unlikely the existence of a much
greater AT for any other important observables.

\section{Measures of Critical Exponents.}\label{EXPONENTS}

Recalling that this model has two different order
parameters~(\ref{MAG},\ref{MAGS}) and, in principle, two different
channels for the correlation lengths, it is compulsory to study
separately both critical behaviors. Therefore, in addition to the
usual exponents ($\nu$,$\beta$,$\gamma$ and $\eta$), we consider the 
analogous for the staggered
channel ($\nus$,$\betas$,$\gammas$ and $\etas$). Although the
most economic scenario requires $\nu=\nus$, in order to consistently
define a single continuous limit, we shall check this assumption.

\subsection{Finite-size scaling.}

Our measures of the critical exponents are based on the FSS ansatz.
Let $\langle O(L,\beta) \rangle$ be the mean value of an operator $O$
measured on a size $L$ lattice, at a coupling value $\beta$, and let
$\xi(L,\beta)$ be a reasonable estimator for the correlation length
on a finite lattice, such as the one shown in eq.~(\ref{XI}). Then, if
$O(\infty,\beta)\sim|\beta-\betac|^{-x_O}$, from the FSS ansatz one
readily obtains~\cite{LIBROFSS}

\begin{equation}
\langle O(L,\beta) \rangle=L^{x_O/\nu}
\left[F_O(\xi(L,\beta)/L) +L^{-\omega}
G_O(\xi(L,\beta)/L) \right]+ \ldots,
\label{FSS}
\end{equation}

where $\omega$ is the universal exponent associated to the leading
corrections-to-scaling. The dots stand for higher order
scaling corrections and $\xi(\infty)^{-\omega}$ terms, negligible in
the critical region. 
The above expression is
straightforwardly generalized to functions of mean values.

We remark that to obtain eq.~(\ref{FSS}) the condition on the
definition of the correlation length is that $\xi(L,\beta)/L$ should
be a smooth monotonous function of $L/\xi(\infty,\beta)$ and that
$x_\xi=\nu$.  That condition is verified by the quantity defined in
eq.~(\ref{XI}) but also by $\kappa L$ (see eq.~(\ref{PSEUDOBINDER}))
and their staggered counterparts. However, the scaling function $G_O$ is
not invariant under a change of the correlation length definition, and
a proper choice can largely reduce it.

Let us emphasize that the fulfillment of the hyperscaling relations is
an {\it a priory} condition of the ansatz.  In fact, if a relation
like (\ref{FSS}) holds, for example, for the magnetization and its
square, one readily obtains 
$\gamma+2\beta=d\nu$. However, we will check the observance of the
hyperscaling relations as a test of consistency.

\subsection{Measuring at the maxima.}

Usually, the application of FSS consists on measuring the mean value of
an observable for different lattice sizes in the \emph{apparent}\/
critical point, where $L/\xi(L,\beta)$ is assumed to be constant, so
the exponent of $L$ is easily extracted, after ensuring that the
corrections-to-scaling are small. Typically one can define two such
points, one is the infinite volume critical point (see
section~\ref{CORRSCAL}), the other is the peak of some quantity like
the connected susceptibility defined as
\begin{equation}
\chi^\mathrm{c}=\chi-VM^2.
\end{equation}

The advantage of measuring at a peak is that the apparent critical
point is self defined, and also that the statistical error is much
smaller at a peak (zero $\beta$ derivative) than when the slope is
large.

\begin{figure}[t]
\epsfig{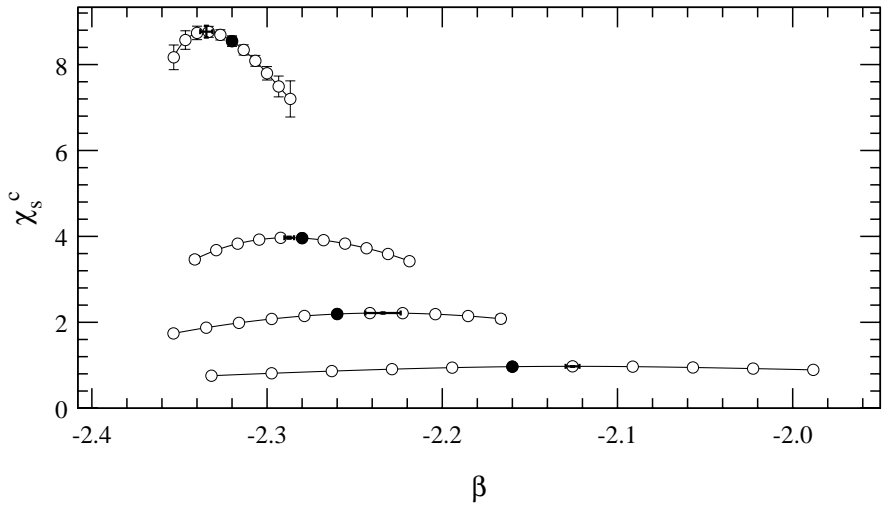}
\caption{Connected staggered susceptibility as a function of lattice
sizes (from top to bottom) $L=24,16,12,8$. The left limit of the
$\beta$ axis corresponds to $\betac$.}
\label{SUSCEPCON}
\end{figure}

In fig.~\ref{SUSCEPCON} we plot the connected staggered susceptibility
for several lattice sizes. The most significant property is the large
shift of the apparent critical point, that makes necessary to perform
simulations at values of the couplings dependent on the lattice size,
as $\betac$ cannot be reached with the spectral density method. The MC
data used in fig.~\ref{SUSCEPCON} correspond to short runs performed
at the coupling values signed by filled symbols.

The shift of the $\beta$ value corresponding to the maximum of the
staggered susceptibility, $\Delta \betac^L$, can be used to measure the
thermal exponent $\nu$ and the critical coupling in the thermodynamic
limit. The expected behavior is
\mbox{$\Delta\betac^L\propto L^{1/\nu}$}. Performing a three parameters
fit, we obtain for $L=8,12,16,24$: \mbox{$\nus=0.83(11)$},
\mbox{$\betac=-2.410(22)$}, with $\chi^2/\mathrm{d.o.f}=0.05/1$.

We can also compute $\etas$ from the value of the susceptibility at
the maximum. 
The fit gives
$\etas=0.035(39)$, with \mbox{$\chi^2/\mathrm{d.o.f.}=0.98/1$}.  The
value for $\etas$ is compatible with those obtained below, but the
errors are 30-times greater (the statistics used is 200-times
smaller).  This method has three important drawbacks compared with the
methods discussed below: a) it is less precise;
b) it needs simulations performed far from the critical point at a
coupling value that depends on the observable to study; and c) large
corrections-to-scaling are expected due to the large distance from the
critical point.

Other quantities, as $\beta$ derivatives of magnetizations or
correlation lengths also present peaks, but they are usually very
broad and noisy what implies a low quality of the obtained critical
exponents. The specific heat does not present a critical divergence as
the $\alpha$ exponent is negative (see subsection~\ref{EXPONENTS2}).

\subsection{The quotient method.}\label{QUOSUBSEC}

Fortunately, one can use another method that does not require a
previous knowledge of the critical point, and which is directly based
on eq.~(\ref{FSS}).  Let $O$ be an observable whose mean value is
measured at the same coupling in two lattices of sizes $L_1$ and $L_2$
respectively.  Using (\ref{FSS}) we can write for their quotient
\begin{equation}
Q_O=\frac{\langle O(L_2,\beta)\rangle}{\langle O(L_1,\beta)\rangle}=
s^{x_O/\nu}\frac{F_O\left(\xi(L_2,\beta)/L_2\right)}
       {F_O\left(\xi(L_1,\beta)/L_1\right)}+O(L^{-\omega}) ,
\end{equation}
where $s=L_2/L_1$. In the case of $O\equiv\xi$, the exponent $x_\xi$
is just $\nu$. Choosing the coupling in such a way that $Q_\xi=s$ we
can directly compute the exponent for any operator as
\begin{equation}
\left.Q_O\right|_{Q_\xi=s}=s^{x_O/\nu}+O(L^{-\omega}).
\label{QO}
\end{equation}
As the exponents are obtained from just two independent simulations,
it is possible to do a very clear statistical analysis.  The use of
the spectral density method avoids the necessity of an {\it a priori}\/
knowledge of the coupling where the quotient of correlation lengths is
the desired.

\begin{figure}[t]
\epsfig{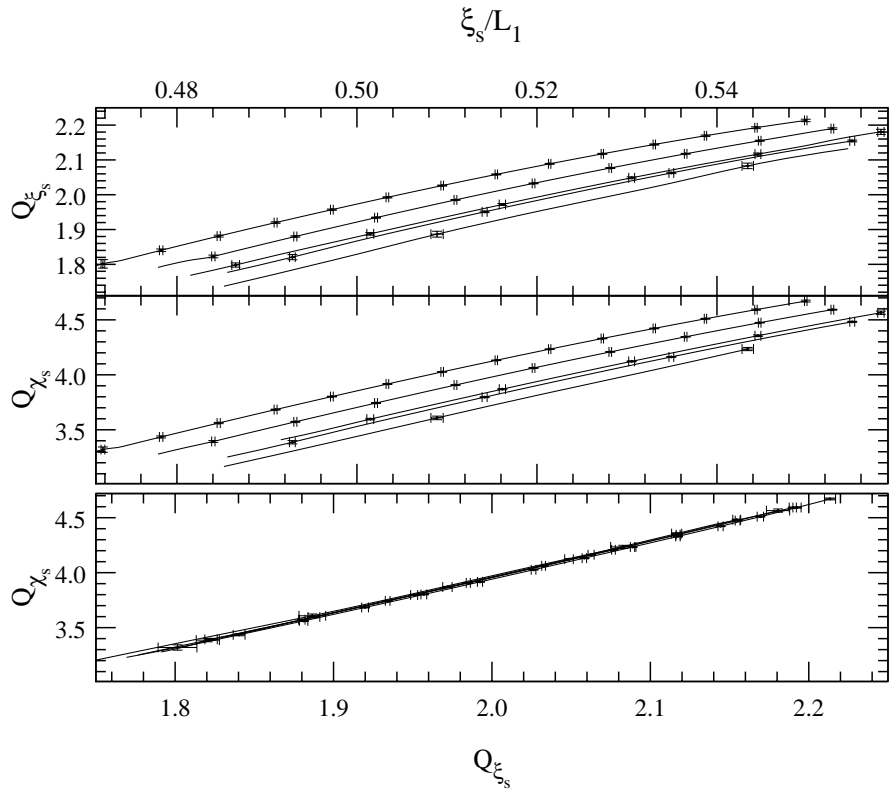}
\caption{Quotient of staggered correlation lengths for $L_2/L_1=2$ as
a function of $\xis(L_1)/L_1$ (upper side). In the middle we plot the
same quantity for the staggered susceptibility.  In both cases the
higher line corresponds to $L_1=6$ with lattice sizes increasing with
decreasing heights.  Finally (lower side) $Q_{\chis}$ as a function of
$Q_{\xis}$ for $L_2/L_1=2$ is plotted.}
\label{QCHISQXIS}
\end{figure}

Let us show with an example how does it work. In fig.~\ref{QCHISQXIS}
(upper part) we plot the quotient between the staggered correlation
lengths for $L_2/L_1=2$ as a function of the correlation length in the
smaller lattice, in lattice size units.  For $L_1$ large we should
obtain a single curve. The deviations correspond to
corrections-to-scaling.  If we plot the quotient for the staggered
susceptibility, we observe similar corrections-to-scaling (see
fig.~\ref{QCHISQXIS}, middle part).  Plotting $Q_{\chi_\stag}$ as a
function of $Q_{\xis}$ the corrections-to-scaling are strongly reduced
even for lattice sizes as small as $L_1=6$ (see fig.~\ref{QCHISQXIS},
lower part).

With this technique, we have studied several operators as the
susceptibility ($x_\chi=\gamma$), the magnetization ($x_M=-\beta$) as
well as their staggered counterparts. To compute the $\nu$ exponent,
we use $\beta$ derivatives of observables. For example,
$x_{\d\xi/\d\beta}=\nu+1$.  In the next two subsections we shall
separately study the determination of $\nu$ and $\eta$ type exponents.

\subsubsection{The exponent $\nu$.}\label{NUEXPSEC}

In the first two columns of table~\ref{NUEXP}, we show the values
obtained for the $\nu$ exponent measuring, at $Q_{\xi_s}=2$, the
$\beta$ derivatives of $\xis$ and $\xi$. The errors have been computed
using the jack-knife method. The corrections-to-scaling are smaller
than the statistical errors, even for lattice sizes as small as $L=6$.

We have repeated the analysis, using the non-staggered correlation
length as the variable. In columns 3 and 4 we write down the obtained
results. These are fully compatible with the previous, but with a
slightly greater statistical error.

\begin{table}[b]
\caption{$\nu$ exponent obtained from a FSS analysis using data from
lattices of sizes $L$ and $2L$. In the first row we show the
correlation length that we ask for duplicate, and in the second one
the operator whose quotient we measure.}
\medskip
\hrule\hrule
\begin{tabular*}{\hsize}{@{\extracolsep{\fill}}rcllcllcll}
    && \multicolumn{2}{c}{$\xis$} 
    && \multicolumn{2}{c}{$\xi  $} 
    && \multicolumn{2}{c}{$\xis$}      \\
\cline{3-4} \cline{6-7} \cline{9-10}
$L$ && \multicolumn{1}{l}{$\!\d\xis/\!\d\beta$} 
    &  \multicolumn{1}{l}{$\!\d\xi /\!\d\beta$}
    && \multicolumn{1}{l}{$\!\d\xis/\!\d\beta$} 
    &  \multicolumn{1}{l}{$\!\d\xi /\!\d\beta$} 
    && \multicolumn{1}{c}{$\!\!\!\!\d \log\! M_\stag/\!\d\beta$} 
    &  \multicolumn{1}{c}{$\!\!\!\!\d \log\! M  /\!\d\beta$} 
\\\hline
6  && .786(6) & .790(6) && .774(14)& .759(16)&&.689(4)  & .670(4)\\
8  && .785(4) & .781(4) && .771(7) & .774(7) &&.717(3)  & .706(2)\\
12 && .789(8) & .782(9) && .781(18)& .781(15)&&.739(9)  & .731(8)\\
16 && .786(6) & .780(5) && .782(8) & .781(6) &&.755(4)  & .751(4)\\
24 && .77(2)  & .77(2)  && .77(3)  & .77(2)  &&.756(16) & .758(13)
\end{tabular*}
\hrule\hrule
\label{NUEXP}
\end{table}

Finally (columns 5 and 6 of table~\ref{NUEXP}) we have alternatively
computed the $\nu$ exponent using the logarithmic derivative of the
magnetizations, which scales as $L^{1/\nu}$. The results have a
smaller statistical error, but they suffer from important
corrections-to-scaling, and it is necessary to compute the
\mbox{$L\to\infty$} extrapolation. This introduces uncertainties
reducing the quality of these numbers as compared with the
previous. Other quantities such as $\beta$-derivatives of Binder's
Cumulants suffer from statistical errors about three times larger, 
and therefore are quite useless for comparison.

We remark that all the determinations of the $\nu$ exponents
(regardless the observables used being staggered or not) are
compatible within errors (at the 1\% level).  One relevant question is
whether summing data from different observables it is possible to
reduce the statistical error, or even if the differences between
columns are significant. As all the used quotients are flat functions
of $Q_\xi$, the main contribution to the error is due to the
observable (not to the correlation length). We have studied the mean
and the difference of the first two columns. The reductions of the
error bars in the first case are negligible (less than a 10\%) because
the data are strongly correlated. In the case of the difference, the
correlation makes the error slightly smaller (about a 10\%) than each
term. The results do not show any significant deviation from zero.

The results for pairs with $s=3/2$ or $s=4/3$, that can be formed with the
simulated lattice sizes, are compatible. 
As our better statistics are in the $L=16,32$ lattices, we only report
the results for $s=2$.

\subsubsection{$\eta$ exponents.}\label{ETAEXPSEC}

The $\eta$ exponent can be obtained from the scaling of the
magnetization or of the susceptibility, using the scaling relations:
$2\beta/\nu=d-2+\eta$ and $\gamma/\nu=2-\eta$.

The results for the $\etas$ exponent are given in table
\ref{ETASEXP}. They have been obtained studying, in the staggered
channel, the susceptibility, the magnetization and the maximum
eigenvalue of the magnetization tensor (that scale with $\gammas$,
$\betas$, and $\betas$ respectively).  Technically, the computation is
somewhat different than in the case of the $\nu$ exponent, since the
quotients change now very fast with $Q_{\xis}$. One has to consider
carefully the statistical correlation of the data. Thus we are able to
measure $\gammas/\nu$ with an accuracy of about a 0.1\%, from which we
can compute $\etas=2-\gammas/\nu$ with a 5\% error.

For the non-staggered channel the results are reported in table
\ref{ETAEXP}. The columns 2, 3 and 4 correspond to the use of $Q_{\xis}$
as the variable. We observe important
corrections-to-scaling. The use of $Q_\xi$ does not improve the
results. 
Using $Q_{\kappa L}$, we
find a quite strong reduction of the corrections-to-scaling.

\begin{table}[b]
\caption{$\etas$ exponent obtained from the FSS
of the staggered susceptibility, magnetization and maximum eigenvalue
of the magnetization, using $\xis$ as the variable.
In all cases, the pairs are of type $(L,2L)$.}
\medskip
\hrule\hrule
\begin{tabular*}{\hsize}{@{\extracolsep{\fill}}rclll}
$L$ && \multicolumn{1}{c}{$\chis$} 
    &  \multicolumn{1}{c}{$M_\stag$}
    &  \multicolumn{1}{c}{$\lambdas^\mathrm{max}$}
\\\hline
6   && 0.0431(10)  & 0.0474(9)    & 0.0480(12) \\
8   && 0.0375(7)   & 0.0409(8)    & 0.0409(9)  \\
12  && 0.0357(17)  & 0.0382(18)   & 0.039(2)   \\
16  && 0.0375(12)  & 0.0395(12)   & 0.0395(13) \\
24  && 0.038(5)    & 0.038(5)     & 0.035(6)   
\end{tabular*}
\hrule\hrule
\label{ETASEXP}
\end{table}

\begin{table}[b]
\caption{$\eta$ exponent obtained from 
the susceptibility, magnetization and maximum eigenvalue
of the magnetization in $(L,2L)$ lattices.
In the first row we indicate the quantity used as the variable.}
\medskip
\tabcolsep=4pt
\hrule\hrule
\begin{tabular*}{\hsize}{@{\extracolsep{\fill}}rclllclllclll}
    && \multicolumn{3}{c}{$\xis$} 
    && \multicolumn{3}{c}{$\kappa L$} \\ 
\cline{3-5} \cline{7-9}
$L$ && \multicolumn{1}{c}{$\chi$} 
    &  \multicolumn{1}{c}{$M$}
    &  \multicolumn{1}{c}{$\lambda^\mathrm{max}$}
    && \multicolumn{1}{c}{$\chi$} 
    &  \multicolumn{1}{c}{$M$}
    &  \multicolumn{1}{c}{$\lambda^\mathrm{max}$}
\\\hline
6 &&1.442(2)&1.447(2)&1.442(3)  &&1.332(6) & 1.332(6) & 1.332(5)  \\
8 &&1.413(2)&1.416(2)&1.4135(18)&&1.332(4) & 1.332(4) & 1.334(4)  \\
12&&1.391(3)&1.393(3)&1.396(4)  &&1.320(13)& 1.321(13)& 1.327(12) \\
16&&1.381(3)&1.383(3)&1.384(3)  &&1.339(5) & 1.340(5) & 1.343(5)  \\
24&&1.362(8)&1.365(9)&1.362(10) &&1.334(18)& 1.334(18)& 1.332(17) 
\end{tabular*}
\hrule\hrule
\label{ETAEXP}
\end{table}

\section{Corrections to scaling.}\label{CORRSCAL}

\begin{figure}[t]
\epsfig{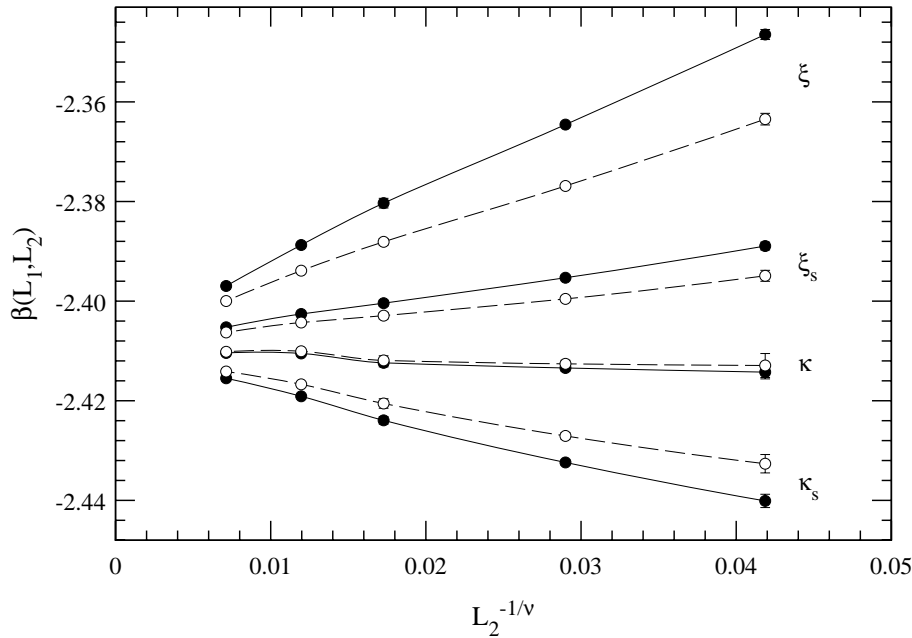}
\caption{Crossing points of scaling functions for pairs $(L_1,L_2)$, as
a function of $L_2^{-1/\nu}$, for $L_1=6$ (solid lines) and $L_1=8$
(dashed line), and $L_2=12,16,24,32,48$.}
\label{CROSS68}
\end{figure}

For the determination of the corrections-to-scaling exponent,
$\omega$, we have used the property shared by different observables
(Binder type parameters, $\kappa$, and correlation lengths divided by
the lattice size) that asymptotically should cross at the critical
point.  It can be shown~\cite{BINDER} that the scaling functions
corresponding to lattice sizes $L_1,L_2$ cross at a $\beta$ value
whose shift from the critical value behaves as
\begin{equation}
\Delta\beta^{L_1,L_2}\propto
\frac{1-s^{-\omega}}{s^{\frac{1}{\nu}}-1}L_1^{-\omega-\frac{1}{\nu}},
\label{SHIFTBETA}
\end{equation}
where $s=L_2/L_1$. 
In fig.~\ref{CROSS68} we plot the values of the
coupling where several scaling functions cross, as a function of
$L_2^{-1/\nu}$. We use along this subsection the value $\nu=0.783$.
 According to (\ref{SHIFTBETA}) if we fix $L_1$, the
asymptotic behavior should be linear, what is well satisfied in the
plot.

\begin{figure}[t]
\epsfig{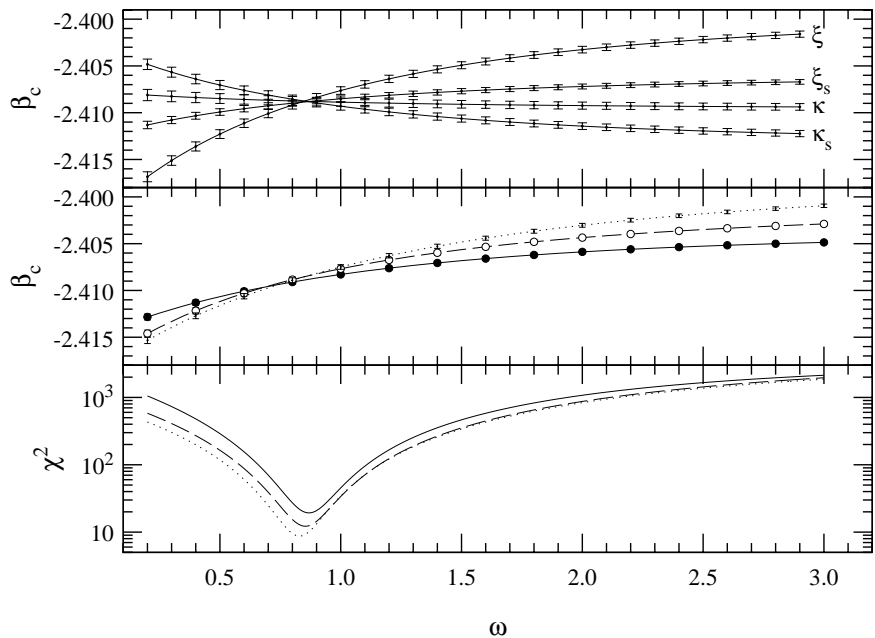}
\caption{Critical coupling extrapolated from the intersection with
$L_1=6$ for $\kappa$, $\kappa_{\stag}$, $\xi/L$ and $\xis/L$ as a
function of the $\omega$ exponent (upper part).
Fits to a single $\betac$ value for $L_1=6$ (solid line), 
$L_1=8$ (dashed line), and $L_2/L_1=2$ (dotted line) as a function of
$\omega$ (middle). In the lower side we show the
values of $\chi^2$ for the latter fits.}
\label{SHIFTFITS}
\end{figure}

We have performed a global fit of the points from each observable to a
functional dependence of the type of eq.~(\ref{SHIFTBETA}) with $L_1=6$.
The good precision obtained for the $\nu$ exponent makes negligible
the uncertainty that comes from this quantity. In fig.~\ref{SHIFTFITS}
(upper part) we plot the values of the extrapolated $\betac$ from independent 
fits for every observable after fixing the value of the
corrections-to-scaling exponent $\omega$.

Studying the variation of the $\chi^2$ function (using the full
covariance matrix) we obtain

\begin{equation}
\begin{array}{lll}
\kappa_\stag:&\betac=-2.4097(19), &\omega=1.05(45),\\
\xi/L:  &\betac=-2.4096(23), &\omega=0.75(20),\\
\xis/L:&\betac=-2.4082(13), &\omega\in(0.6,1.9).
\end{array}
\label{FITSINDEP}
\end{equation}
The corrections-to-scaling in the $\kappa$ case are so small that it
is not possible to measure the parameters. Repeating the fits with
$L_1=8$ we obtain very similar results.

From these fits we conclude that all the extrapolated values coincide
up to $\Delta\beta=0.002$. Notice that this implies that both
correlation lengths diverge at values of $\beta$ which are at most
$0.002$ apart. We can go one step further under the hypothesis that
the extrapolated values must be the same, which imply that the
infinite volume propagator has two separated poles in momentum space,
one at the origin, the other at $(\pi/a,\pi/a,\pi/a)$. A $\chi^2$
minimization of all the (strongly correlated) data gives
(see fig.~\ref{SHIFTFITS}, middle and lower parts)
\begin{equation}
\betac=-2.4088(3), \omega=0.87(3), \chi^2/\mathrm{d.o.f.}=19.3/14 .
\label{FITBETAC6}
\end{equation}
We have repeated the fits discarding the $L_1=6$ data, and measuring
the crossing with the $L_1=8$ curves. We obtain
\begin{equation}
\betac=-2.4085(3), \omega=0.86(4), \chi^2/\mathrm{d.o.f.}=12.2/14,
\label{FITBETAC8}
\end{equation}
that are fully compatible with a very slight increase of the
statistical errors.  This gives us confidence in the smallness of the
higher order corrections-to-scaling.

A further check can be done studying the crossing between lattice
sizes with $s=2$. The functional form of the fit is different now, as
$s$ in eq.~(\ref{SHIFTBETA}) is fixed. We obtain, discarding the $L=6$ data,

\begin{equation}
\betac=-2.4087(4), \omega^*=0.83(4), \chi^2/\mathrm{d.o.f.}=8.7/10.
\label{FITBETAC}
\end{equation}

where $\omega^*$ is defined as $\omega+\frac{1}{\nu}-(0.783)^{-1}$.
 
We will use in the next subsection these results to study the infinite
volume extrapolation of the critical exponents.  On the other hand,
with our $\betac$ estimation, we can use again eq.~(\ref{FSS}) to obtain
another estimation of critical exponents. This shall be done in
subsection~\ref{EXPONENTS2}.

\subsection{The infinite volume extrapolation of the exponents.}

The FSS analysis has the nice feature of using the finite-size effects
to measure quantities as important as critical exponents.  However,
this does not guarantee the absence of systematic errors coming from
finite-size effects, that appear as corrections-to-scaling terms.

Let $x$ be the critical exponent under study, and let us suppose that we
need to take care of just the first term on the corrections-to-scaling
series, then we have from eq.~(\ref{QO}),
\begin{equation}
\left.{\frac{x}{\nu}}\right|_{(2L,L)}=\frac{x}{\nu}+aL^{-\omega}+ \ldots,
\end{equation}
$a$ being a constant.

We have carried out a fit, to the above functional form, for the
numbers obtained in subsections~\ref{NUEXPSEC} and~\ref{ETAEXPSEC}.
We exclude the $L=6$ data, and take for the universal exponent
$\omega$ the value obtained previously ($\omega=0.85(4)$).

We have found two very different situations. Some estimators present a
non measurable size dependence (first four columns of
table~\ref{NUEXP}, the whole table~\ref{ETASEXP}, and the last three
columns in table~\ref{ETAEXP}). For them, we find no dependence on
$\omega$ (when it moves within its error bars) of the extrapolated
exponent, that has a value almost equal to the most relevant number in
the tables (the $L_1=16$, $L_2=32$ pair), but the errors are doubled.

We also have found estimators with a quite measurable size dependence,
namely the last two columns of table~\ref{NUEXP} (exponent $\nu$), and
the first three columns of table~\ref{ETAEXP} (exponent $\eta$). 
The extrapolated values are
\begin{equation}
\begin{array}{c}
\nu_5=0.800(9),\ \nu_6=0.804(8),\\
\eta_1=1.338(6),\ \eta_2=1.339(6),\ \eta_3=1.346(6),
\end{array}
\end{equation}
where the subindices refer to columns in tables~\ref{NUEXP}
and~\ref{ETAEXP}.  We find here an $\omega$-dependence of the
extrapolated exponents of one half of the quoted statistical error.
While the extrapolation for the $\eta$ exponent is fully consistent
with the quoted numbers in the last three columns of
table~\ref{ETAEXP}, the $\nu$ we get looks too big. Although both
estimations are within two standard deviations, higher order
corrections-to-scaling are apparent by plotting the last columns of
table~\ref{NUEXP} as a function of $L^{-\omega}$.

As a final result for the exponents, we choose operators that give
stable values with the lattice size, taking the measure for the
$(16,32)$ pair;  the number in the second parenthesis shows the error
increase due to the extrapolation that can be taken as an
estimation of the error involved in this procedure:

\begin{equation}
\begin{array}{lcllcllll}
\nu   &=&0.783(5)(6),\\ 
\etas &=&0.0380(12)(14),
      &\quad(\betas&=&0.406(3)(3),\ \gammas&=&1.536(10)(12)&),\\
\eta  &=&1.339(5)(5)
      &\quad(\beta &=&0.916(6)(7),\ \gamma &=&0.518(6)(7)&).
\end{array}\label{NUFINAL}
\end{equation}

For comparison, we quote the MC results for $\nu$ obtained for the three
dimensional ferromagnetic $\Orth3$~\cite{O33D} and $\Orth4$~\cite{O43D} models:
\begin{equation}
\begin{array}{rcl}
\nu(\Orth3) &=& 0.704(6),\\
\nu(\Orth4) &=& 0.7479(90).
\end{array}
\label{EXPOON}
\end{equation}
Our results for the $\RP2$ model are completely incompatible to those
of the $\Orth3$ one. Regarding the $\Orth4$ model, although the
differences are small, the value for the $\nu$ exponent seems also
incompatible.

\subsection{Critical exponents measuring at ${\mathbf{\betac}}$.}
\label{EXPONENTS2}

Taking the values $\betac=-2.4086(4)$ and $\omega=0.85(4)$ we will
repeat the measurements of the critical exponents using the measures
of the usual operators at a fixed coupling.  As in the determination
of $\betac$ we use all the MC data, it is difficult to perform a
correct computation of the statistical errors taking into account all
correlations. In addition, there are several sources of systematic
errors, consequently the evaluation of the errors on the exponent is
harder.

The functional form we fit is $AL^{x/\nu}(1+bL^{-\omega})$, where
$\omega$ is fixed.  We present the results for the computed exponents
in table~\ref{EXPOBETAC}.  In the first column we show the minimum
value of $L$ that we include in the fit. We remark that the inclusion
of the $L^{-\omega}$ term is essential to get a correct fit.  To
estimate the errors, we have extrapolated at our measured $\betac$ and
at one standard deviation, adding the difference to the statistical
error in the extrapolation. The $\omega$ error contribution 
is negligible.

\begin{table}[b]
\caption{Critical exponents obtained by measuring at $\betac$. $L$ stands
for the minimum value of the lattice size we have used for the fit.
The contribution of the error in $\betac$ and $\omega$ is included.}
\medskip
\hrule\hrule
\tabcolsep 4pt
\begin{tabular*}{\hsize}{@{\extracolsep{\fill}}rcllcllcll}
 &&  \multicolumn{2}{c}{$\nu$} && \multicolumn{2}{c}{$\etas$} 
 && \multicolumn{2}{c}{$\eta$}       \\
\cline{3-4} \cline{6-7} \cline{9-10}
\multicolumn{1}{c}{$L$}&&\multicolumn{1}{c}{$d\xis/d\beta$}
    & \multicolumn{1}{c}{$d\xi/d\beta$}
    && \multicolumn{1}{c}{$\chis$}      
    & \multicolumn{1}{c}{$M_\stag$}
    && \multicolumn{1}{c}{$\chi$}
    & \multicolumn{1}{c}{$M$}\\\hline
6  &&0.783(9) &0.761(9) &&0.030(7) &0.030(7) &&1.324(12)& 1.323(13)  \\
8  &&0.784(15)&0.771(14)&&0.023(9) &0.023(10)&&1.316(16)& 1.316(16)  \\
12 &&0.79(3)  &0.79(3)  &&0.011(16)&0.010(17)&&1.292(27)& 1.291(27)  \\
16 &&0.75(7)  &0.82(8)  &&0.02(3)  &0.02(3)  &&1.31(5)  & 1.31(5)    
\end{tabular*}
\hrule\hrule
\label{EXPOBETAC}
\end{table}

For the $\eta$ exponents, as the slope of $M$ and $\chi$ as a function
of $\beta$ is large at the critical point, the errors in the
determination of $\betac$ affect strongly that of $\eta$. This error
is, in part, overestimated since we do not take into account the
statistical correlation between the measures used to determine
$\betac$ and the magnetization operators. This, albeit complex, could
have been done. However, as the $\betac$ value has been obtained after
a somewhat elaborated procedure, the influence of systematic errors is
difficult to quantify. Consequently we consider more solid the
determination obtained from the quotient method.

\begin{figure}[t]
\epsfig{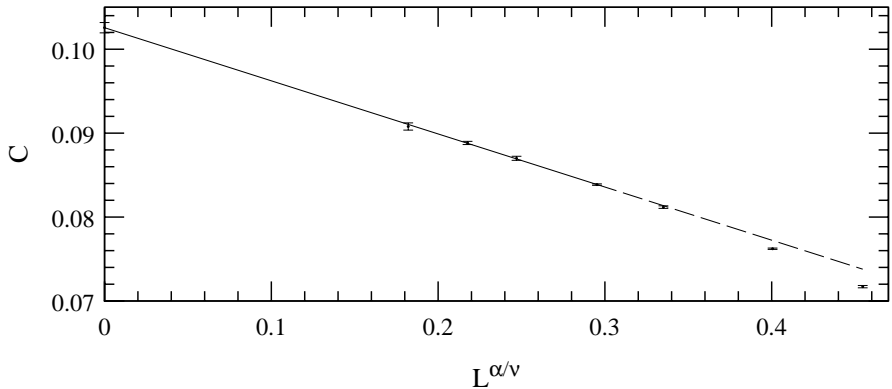}
\caption{Specific heat at the critical point for all lattice sizes as
a function of $L^{\alpha/\nu}$. The linear fit has been done for
$L\geq 16$}
\label{CALESP}
\end{figure}

From the hyperscaling relation $\alpha=2-d\nu$, using
(\ref{NUFINAL}), we have \mbox{$\alpha/\nu=-0.44(2)$} and therefore, no
divergence is to be seen for the specific heat. We could nevertheless
stick to the FSS prediction. In fact in fig.~\ref{CALESP} we show the
specific heat at the critical point as a function of
$L^{\alpha/\nu}$. 
Strong scaling corrections are apparent, but the overall linear
behavior is well satisfied.

\section{Vacuum structure in the critical region.}\label{VACUUM}

The antiferromagnetic broken phase is characterized by a $L\to\infty$
nonzero value for the magnetizations $M$ and $M_\stag$. However,
studying the different terms of the tensors we can obtain a lot of
information about the structure of the broken phase.  We shall
consider the behavior of several operators constructed from the
magnetization tensors.  As the staggered magnetization behaves as
$t^{\betas}$ ($\betas\sim 0.4$) and the magnetization as $t^{\beta}$
($\beta\sim 0.9$) the former is dominant for $t$ small, $t$ being the
reduced temperature.

The first question we address is the breakdown of the even-odd
symmetry.
Near to the critical point it
is very difficult to reach an asymptotic ($L\to\infty$) behavior that
allows us to know if that symmetry is broken, but instead we can use
FSS arguments to find the answer.

Let us consider the tensors product trace 

\begin{equation}
\tr\mathbf{M}_\stag\mathbf{M}=\frac{1}{4}
(\tr\mathbf{M}^2_\mathrm{e}-\tr\mathbf{M}^2_\mathrm{o}),
\label{PRODMMS}
\end{equation}

where $\mathbf{M}_\mathrm{e}=2/V\sum_{i\ \mathrm{even}}\mathbf{T}_i$
and analogously for the odd sublattice. If the even-odd symmetry were
not broken, the $\mathbf{M}_\stag\mathbf{M}$ tensor should go to zero
as $1/V=L^{-3}$, while the LHS of eq.~(\ref{PRODMMS}) is bounded by a
$L^{-(\betas+\beta)/\nu}$ behavior.  In practice we have studied the
quantity $A=\langle(\tr\mathbf{M}_\stag\mathbf{M})^2\rangle$.

We have used again the technique described in subsection \ref{QUOSUBSEC}
to determine the critical exponent.
We obtain, after an $L\to\infty$ extrapolation the value
\begin{equation}
\frac{x_A}{\nu}=-3.389(15),
\end{equation}
which is compatible with 
$-2(\beta+\betas)/\nu=-3.377(13)$, and discards the $-6$ value.
We thus conclude that the even-odd symmetry breakdown does occur.

The naive picture that comes from the previous conclusion is that one
of the sublattices is plane aligned and the other fully aligned, and
the $\Orth{2}$ global symmetry is preserved. To check this we have
studied the conmutator of the $\mathbf{M}_\stag$ and $\mathbf{M}$
tensors.  If there were a remaining $\Orth{2}$ symmetry, both tensors
could be simultaneously diagonalized and the conmutator should
vanish. In a finite lattice, according to FSS it should behave as
$L^{-d/2-\betas/\nu}$.  Our results show that the exponent for
$B=\langle\tr[\mathbf{M}_\stag,\mathbf{M}]
[\mathbf{M}_\stag,\mathbf{M}]^\dagger\rangle$ has strong
corrections-to-scaling, but  after a $L\to\infty$ extrapolation we obtain
\begin{equation}
\frac{x_B}{\nu}=-3.406(11).
\end{equation}
Due to these corrections the error bar is probably underestimated. We observe 
that is again very different from $-4.04(2)$ and compatible with 
$-2(\beta+\betas)/\nu$.
The consequences of these results are that the $\Orth{2}$ symmetry must
be broken, and that the alignment directions of both sublattices
are not orthogonal.

\section{Numerical results at low temperature.}\label{LOWTEMP}

\begin{figure}[t]
\epsfig{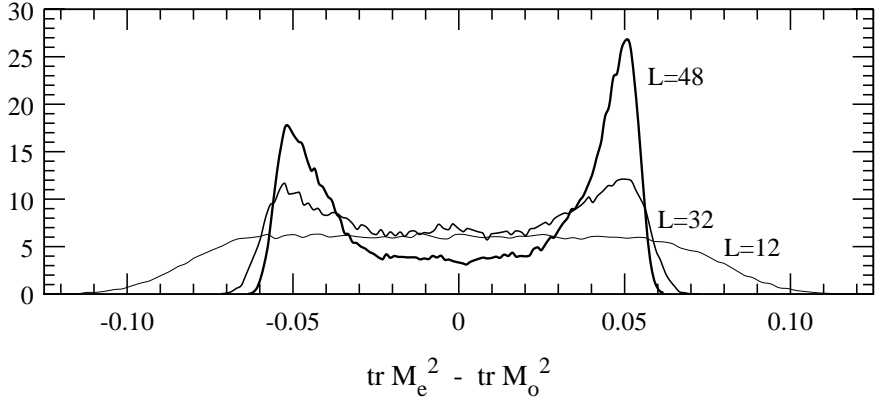}
\caption{Histogram of $\tr\mathbf M_\mathrm{e}^2-\tr\mathbf
M_\mathrm{o}^2$ at $\beta=-3.0$ for several lattice sizes.}
\label{trMMs}
\end{figure}

We present in this section some numerical results obtained from
simulations in the low temperature ($\beta<\betac$) region. The very
first thing to establish is the existence of a low temperature phase,
where both order parameters are not zero, and where the symmetry
between the even and odd sublattices is broken, as discussed in the
previous section. We can answer both questions computing an histogram
of $\tr(\mathbf{M}\mathbf{M}_\stag)$. This is shown in
fig.~\ref{trMMs} for $\beta=-3.0$. We see a clearly nonzero
thermodynamic limit for $\tr (\mathbf{M}\mathbf{M}_\stag)$.
For $\beta<-3$, the double peak structure is even clearer.

\begin{figure}[t]
\epsfig{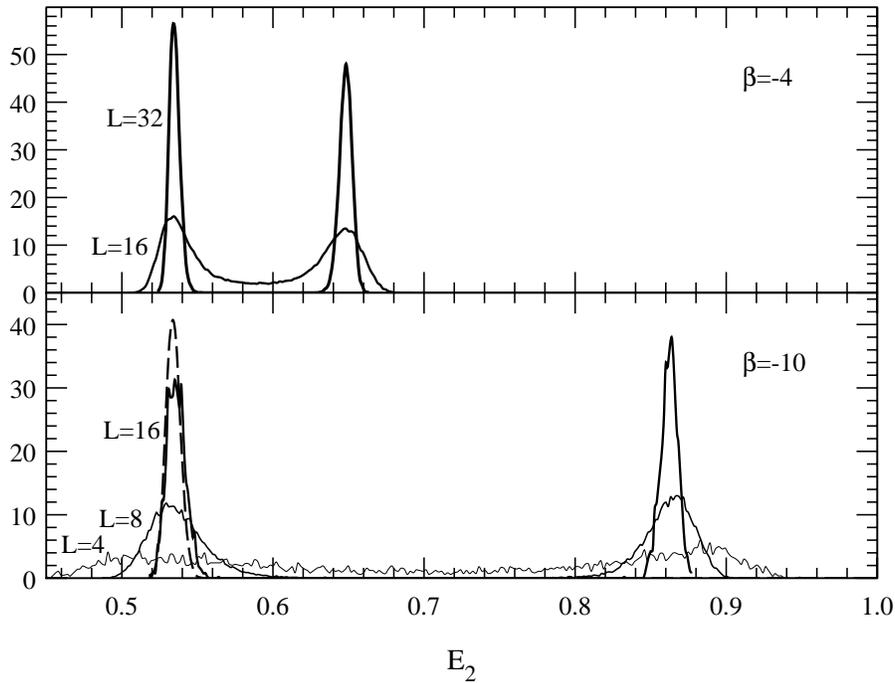}
\caption{Histogram of the next-nearest neighbors energy for the even
and odd sublattices. The dashed line corresponds to the effective
(flat) low temperature model for $L=16$.}
\label{O2LOCAL}
\end{figure}

We have next studied the second-neighbor energy for the odd and even
sublattices independently. In fig.~\ref{O2LOCAL} we plot the
histograms of the sublattice energy. We recall that for
$\beta=-\infty$, $E_2=1$ for one of the sublattices and $E_2=0.5$ for
the other.  We observe a clear double peak structure at $\beta=-4$
(upper side).  At $\beta=-10$ (lower side) the double peak structure
is already clear even for lattice sizes as small as $L=4$ and for
$L\ge 8$ the peaks are completely split. The dotted line
corresponds to the flat model (subsection~\ref{FLATMODEL}), this means
that for this $\beta$ value the results are almost asymptotic
($\beta\to-\infty$) for the left peak.

It is very interesting to observe the evolution of the second-neighbor
energy for the less aligned sublattice.  At $-\beta$ very large, the
field variables are almost in a S$^1$ subset of S$^2$, so, we can look
at it as an $\RP1$ model in a face centered cubic (fcc) lattice. This
model (that is equivalent to an $\Orth{2}$ sigma model) has a second
order phase transition. For a rough comparison we have measured the
energy $\langle \cos^2(\varphi_i-\varphi_j)\rangle$ for the fcc
$\RP1$ model at the critical point, obtaining $E_2=0.62(1)$.  

In
the $\RP{2}$ model, the fluctuations produce an effective coupling
between second-neighbors of the less aligned sublattice.
For $-\beta$ very large, we can use the {\em flat} model defined by
(\ref{FLATHAMIL}). We obtain $E_2\approx 0.53$ for the less aligned
sublattice (see fig.~\ref{O2LOCAL}). This value is very small, so the
effective coupling is too weak to produce an $\Orth{2}$ ordering.

For increasing values of $\beta$, the fluctuations off the plane
reduce the $E_2$ value although
$\langle\cos^2(\varphi_i-\varphi_j)\rangle$ increases.  This effect
can be taken into account in a crude approximation as follows.  Let us
write the spin variable as in eq.~(\ref{VFLAT}). The second-neighbor
energy takes the form
\begin{equation}
\begin{array}{rcl}
\langle (\bbox v_i \cdot \bbox v_j)^2 \rangle&=
&\langle (1-(v_i^z)^2)(1-(v_j^z)^2)\cos^2(\varphi_i-\varphi_j)\rangle
+\langle (v_i^zv_j^z)^2\rangle\\
&&+\langle v_i^zv_j^z\sqrt{1-(v_i^z)^2}\sqrt{1-(v_j^z)^2}
 \cos(\varphi_i-\varphi_j)\rangle.
\end{array}
\end{equation}
As a first approximation, we can suppose that the $z$ components of
the neighbors are uncorrelated, and similarly for the $z$ components
with the angles. Thus we can factorize the mean values obtaining
\begin{equation}
E^{\RP1}\equiv\langle \cos^2(\varphi_i-\varphi_j)\rangle\approx
\frac{E_2
-\lambda_\mathrm{min}^2}{(1-\lambda_\mathrm{min})^2},
\label{COSENOESTIM}
\end{equation}
where $\lambda_\mathrm{min}=\langle (v_i^z)^2\rangle$ is the minimal
eigenvalue of the magnetization tensor of the less aligned sublattice.

The numerical results obtained for several $\beta$ values are

\begin{equation}
\begin{array}{ll}
\beta=-10,     &E^{\RP1}=0.53(1),\\
\beta=-4,      &E^{\RP1}=0.61(1),\\
\beta=-3,      &E^{\RP1}=0.65(1),\\
\beta=-2.41,   &E^{\RP1}=0.798(2).
\end{array}
\label{COSENO}
\end{equation}

We observe a fast increase of that quantity when approaching to the
critical point.  In particular, the critical value for the $\Orth{2}$
model is reached at $\beta=-4.0$. This argument, although qualitative,
supports the idea that the breakdown of the $\Orth{2}$ symmetry
observed at the critical point can be explained with the induced
interaction of the more ordered sublattice on the less ordered one.

\begin{figure}[t]
\epsfig{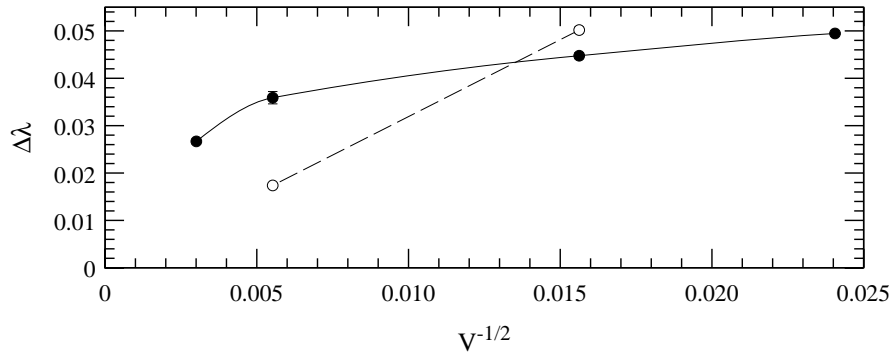}
\caption{Difference of the two smaller eigenvalues of $\mathbf{M}$ as
a function of $V^{-\frac{1}{2}}$
for $L=16,32$ at $\beta=-4.0$ (dashed
line), and for $L=12,16,32,48$ at $\beta=-3.0$ (solid
line).}\label{L2L3}
\end{figure}

We expect that the breakdown of the $\Orth{2}$ symmetry occurs at
$\beta<\betac$ since $E^{\RP1}$ becomes very large at
$\betac$. However, we have been unable to obtain a clear
thermodynamic limit at $\beta<\betac$ for an $\Orth2$ broken phase.
Let us state that for an $\Orth2$ symmetric phase, we expect to find a
diminishing difference of the two smaller eigenvalues of the global
magnetization as $L^{-\frac{d}{2}}$.
In fig.~\ref{L2L3}, we plot this difference as a function of
$L^{-\frac{d}{2}}$, for two $\beta$ values.  At $\beta=-4$, the
analysis for $L\le32$ lattices shows an $\Orth2$ symmetric
thermodynamic limit.  At $\beta=-3$, the same analysis seems to
indicate an $\Orth2$ non-symmetric behavior for $L\le32$, but the
$L=48$ data points to a restoration of the symmetry. Finding a clear
non symmetric limit would imply to simulate nearer to $\betac$ in much
larger lattices than those we have used in this work.

There is an alternative procedure that we think easier and more
conclusive that consists in the introduction of a second-neighbor
coupling which explicitly breaks the $\Orth2$ symmetry
\begin{equation}
\mathcal{S}_2= \beta_2 \sum_{\ll i,j\gg} (\bbox{v}_i\cdot\bbox{v}_j)^2 .
\end{equation}

Let us discuss some regions of the $(\beta,\beta_2)$ plane. For
$\beta=-\infty$, the line $\beta_2$ corresponds to a fcc $\RP1$
($\equiv \Orth2$) model that presents a second-order phase transition
at $\beta_2=0.415(1)$.  At $-\beta$ large but finite, we expect a second
order line, in the same universality class, almost parallel to the
$\beta$ axis. As the fluctuations of the more aligned sublattice help
in the $\Orth2$ ordering, this line should get closer to the $\beta$ axis
and presumably crosses it before $\betac$.
We project to study this enlarged parameter space shortly.

\section{Conclusions and outlook.}\label{CONCLUS}

We have found very interesting properties in the AF
sector of the $\RP{2}$ spin model in three dimensions. It presents two
magnetization operators with different critical properties, what is
related with an odd property of the vacuum in the critical region: a
complete breakdown of the global $\Orth{3}$ symmetry of the action. We
have accurately measured the critical exponents, using a modest amount
of computer time, by means of a FSS method that performs very good
even in the presence of important corrections-to-scaling.  The
critical exponents obtained are clearly different from those of the
ferromagnetic $\Orth{3}$ model, and hardly compatible with the ones
reported for $\Orth4$.

We project to extend these results to several related models, with
generalizations such as, for instance, the inclusion of
second-neighbor interactions or of vectorial interactions.  Work is
already in progress~\cite{RP2D4} on the four dimensional version of
this model. We are particularly interested in the relevance,
in $d=4$, of the properties found here. In addition, it can
enlighten the mechanism by which the AF transition seems to belong to
a new Universality Class in $d=3$.

\section*{Acknowledgements}
We are indebted to A. Sokal for many suggestions.  We also thank
J.L. Alonso, J.M. Carmona, L.A. Ibort, J.J. Ruiz-Lorenzo and
A. Taranc\'on for discussions. This work has been partially supported
by CICyT AEN93-0604-C03-02 and AEN95-1284-E.

\newpage

\appendix

\section{Details of the Mean-Field computation.}\label{MFAPPEN}

To compute the function ${\mathcal G}_\rho$ defined in (\ref{DES})
we start from the Hamiltonians written in terms of the tensorial products
$\bbox \tau_i=\bbox {v}_i \otimes \bbox {v}_i$,

\begin{equation}
{\mathcal H}=\sum_{<i,j>} \tr \bbox{\tau}_i \bbox{\tau}_j,\\
\end{equation}
\begin{equation}
{\mathcal H}_\rho= -\sum_{i\ \mathrm{even}}\tr\bbox{\tau}_i
\mathbf{Q}_{\mathrm e} -\sum_{i\ \mathrm{odd}}\tr\bbox{\tau}_i
\mathbf{Q}_{\mathrm o},
\end{equation}
where, given the antiferromagnetic character of our model, it is rather
natural to use a (tensorial) mean field for the odd sublattice, and an
independent one for the even sites. In this approximation the
partition function factorizes as

\begin{equation}
{\mathcal Z}_\rho=z_{\mathrm e}^{V/2}z_{\mathrm o}^{V/2},\quad
z_\mathrm{e,o}=\int_{\mathrm S^2}\d \bbox v 
\exp(\frac{1}{T}\tr  \bbox \tau \mathbf Q_\mathrm{e,o}), 
\end{equation}
and the mean value of $\bbox \tau$ can be written as
\begin{equation}
\langle\bbox{\tau}_\mathrm{e,o}\rangle_{\rho}
=\mathbf M_{\mathrm e,o}(\bbox Q_{\mathrm e,o})=(z_\mathrm{e,o})^{-1}
\int_{\mathrm S^2}\d \bbox v \ \bbox\tau
\exp(\frac{1}{T}\tr \bbox \tau \mathbf  Q_\mathrm{e,o}).
\label{MDEF}
\end{equation}

We thus obtain the quantity to minimize, $\mathcal{G}_\rho$, as a
function of $\bbox Q_\mathrm{e},\bbox Q_\mathrm{o}$, and
the temperature $T$, 
\begin{equation}
\frac{2}{V}{\mathcal G}_\rho=
-T\log(z_\mathrm{e}z_\mathrm{o})
-2d\,\tr \mathbf{M}_\mathrm{e}\mathbf{M}_\mathrm{o}
+\tr\mathbf{ M_\mathrm{e} Q_\mathrm{e}}
+\tr\mathbf{ M_\mathrm{o} Q_\mathrm{o}}.
\label{GRHO}
\end{equation}

It is easy to show that ${\mathcal G}_{\rho}$ is invariant if we add
to $\mathbf{Q}_\mathrm{e}$ ($\mathbf{Q}_\mathrm{o}$) an arbitrary
multiple of the identity. Thus, we can shift to zero, for each field,
one of its eigenvalues.

Minimizing general tensorial fields would be very complex, so we
shall restrict ourselves to the case
$[\mathbf{Q}_\mathrm{e},\mathbf{Q}_\mathrm{o}]=0$, where 
they can be simultaneously diagonalized.  From now on, we shall
work in a basis where both fields are diagonal. Let us remark that an
$\Orth2$-symmetric vacuum or a non-symmetric one, can be considered in
this class. We shall write $\mathbf{Q}_\mathrm{e}$,
($\mathbf{Q}_\mathrm{o}$) as the diagonal matrix with eigenvalues
$(q_1,q_2,0)$ ($(q_3,q_4,0)$). Consequently, $\mathbf{M}_\mathrm{e,o}$
are trace-one diagonal matrices, with nonnegative eigenvalues.

The values of the $q_i$ parameters that minimize $\mathcal G_\rho$
satisfy the equations
\begin{equation}
0=\tr \left((2d\/
\mathbf M_{\mathrm e}-\mathbf Q_{\mathrm o})
\frac{\partial \mathbf M_{\mathrm o}} {\partial q_i}\right),\ i=1,2,
\label{MFE}
\end{equation}
and analogously for $i=3,4$ interchanging odd by even.

It is easy to check that $\mathbf Q_{\mathrm{e,o}}=0$ is a solution of
(\ref{MFE}) for all values of $T$.  Nevertheless, this could
correspond to a maximum as well as to a minimum of ${\mathcal
G}_{\rho}$.  In fact, one would expect this to be the minimum for
large $T$, which would turn into a maximum for $T$ small enough.

Eqs.~(\ref{MFE}) are a set of quite complex nonlinear coupled
equations.  To go further, let us assume that the system undergoes a
continuous transition at a critical point $\Tc$. We can thus
linearize them as $0=\sum_{i=1}^4 A_{ji}(T)q_i$, in order to find
a minimum close to $\mathbf Q_\mathrm{e,o}=0$.  To have a non trivial
solution, we require $\det A(T)=0$, which happens at
\begin{equation}
\betac=-\frac{1}{T}=-\frac{15}{4d}\ .
\end{equation}
This nontrivial minima should be in the directions in $q$-space
pointed by the vectors belonging to the kernel of the matrix $A$, that
is spanned by the vectors $(1,0,-1,0)$ and $(0,1,0-1)$. We shall then
restrict ourselves to the subspace \mbox{$q_1=-q_3$},
\mbox{$q_2=-q_4$}.  Notice that there are three directions in this
plane, where the ordered phase keeps an $\Orth2$ symmetry:
$(1,1,-1,-1)$, $(1,0,-1,0)$, and $(0,1,0,-1)$.  Expanding
eq.~(\ref{MDEF}) in powers of $q_i$ it is straightforward to see that
the linear terms are opposite for $\mathbf M_{\mathrm e}$ and $\mathbf
M_{\mathrm o}$ in this subspace, so, at first nonzero order
\begin{equation}
\begin{array}{cclcl}
\mathbf M_{\mathrm s}&=&\frac{1}{2}
(\mathbf M_{\mathrm e}-\mathbf M_{\mathrm o})&\propto &q_i,\\
\mathbf M&=&\frac{1}{2}
(\mathbf M_{\mathrm e}+\mathbf M_{\mathrm o})-\frac{1}{3}\mathbf 1
&\propto &q_i^2,\\
E-\frac{1}{9}&=&\tr \mathbf M_{\mathrm e}\mathbf M_{\mathrm o}-\frac{1}{9}
&\propto &q_i^2.
\end{array}
\end{equation}

To calculate the dependence of $q_i$ on $T$
we need to consider higher order terms in ${\mathcal G}_{\rho}$.
We obtain, using polar coordinates in the $(q_1,q_2)$ plane
\begin{equation}
{\mathcal G}_{\rho}=a+b(T-\Tc)s^2+c(T)s^4+O(s^6),
\label{GRHOS}
\end{equation}
where $s=r\sqrt{2-\sin 2\theta}$, and $b,c(T)>0$.  At leading order we
also find \mbox{$\tr \mathbf M_{\mathrm s}^2\propto s^2$}. For
$T>\Tc$, $s=0$ is a minimum of ${\mathcal G}_{\rho}$, while
when $T<\Tc$, $s=0$ is a maximum and the real minimum is at
$s_\mathrm{min}\propto (\Tc-T)^{1/2}$.  Therefore, the mean
field prediction for the magnetic and specific heat critical exponents is
\begin{equation}
\beta_\mathrm{s}=1/2,\quad \beta=1,\quad \alpha=0.
\end{equation}

Up to this order, the symmetry of the ordered phase cannot be
determined since there are minima for $\mathcal G_\rho$ along all 
directions in the $(q_1,q_2)$ plane.  The sixth order term does break
this degeneracy but favors an $\Orth2$ symmetric ordered phase.  This
is in contradiction with our numerical data just after the transition.

\section{Classical continuum limit.}

We shall derive the naive classical continuum limit for the $\RP2$
antiferromagnetic model. As we are interested in an $\Orth2$ broken
phase, for consistency we should add to the action (\ref{HAMILTONIAN})
a ferromagnetic second-neighbor term that, as we shall see, does not
change the continuum functional form.

Let us first state that any integral over the S$^2$ sphere can be
written as an integral over the $\SO3$ group. If $\bbox u$ is an
arbitrary element of S$^2$, for any function $f$,
\begin{equation}
\int_{\mathrm S^2} \d\bbox v f(\bbox v)=\int_{\SO3} \d R f(R \bbox u),
\end{equation}
where $\d\bbox v$ is the invariant measure over the sphere and 
$\d R$ the Haar measure over the $\SO3$ group, both normalized.
This result is easily generalized to a multiple integral. In
particular the partition function can be expressed as
\begin{equation}
{\mathcal Z}\equiv\int_{S^2}\left(\prod_i \d\bbox v_i\right) 
\e^{-{\mathcal S}(\sbbox v_1,\ldots,\sbbox v_N)}
=\int_{\SO3}\left(\prod_i \d R_i\right)
\e^{-{\mathcal S}(R_1\sbbox u_1,\ldots,R_N\sbbox u_N)}.
\end{equation}
The vectors $\{\bbox u_1,\ldots,\bbox u_N\}$ should be chosen for
$R_i$ to be a smooth function of the spatial position in the
$\beta\to-\infty$ limit, with the second-neighbor ferromagnetic
coupling large enough to ensure the $\Orth2$ breakdown.  For a
selection of the type
\begin{equation}
\bbox u_i=\left\{\begin{array}{ll}
                \bbox u_\mathrm{e},&i\ \mathrm{even},\\
                \bbox u_\mathrm{o},&i\ \mathrm{odd},
                \end{array}
          \right.
\end{equation}
the action can be written, in terms of the $R$ variables, as
\begin{equation}
\begin{array}{rl}
{\mathcal S}&=
\displaystyle 
-\beta\sum_{i\ \mathrm{even}}\sum_{q=\pm 1}\sum_{\mu=1,2,3}
\left[(R_i\bbox u_\mathrm{e})\cdot(R_{i+q\mu} 
        \bbox u_\mathrm{o})\right]^2\\
&\displaystyle 
+\frac{\beta'}{2}\sum_{i\ \mathrm{even}}\sum_{q,p=\pm 1}
         \sum_{\stackrel{\mu,\nu=1,2,3}{\mu\ne\nu}} 
\left[(R_i\bbox u_\mathrm{e})\cdot(R_{i+q\mu+p\nu} 
        \bbox u_\mathrm{e})\right]^2
+\left(\mathrm{even}\leftrightarrow\mathrm{odd}\right).
\end{array}
\end{equation}

We need to take $\bbox u_\mathrm{e} \cdot \bbox u_\mathrm{o} = 0$ in
order $R$ to be smooth.  Thus we can substitute $R$ by its Taylor
expansion, and the action becomes at leading order
\begin{equation}
\begin{array}{rcl}
{\mathcal S}&\approx&\mathrm{constant}
+\displaystyle
\frac{\beta}{a}\int_{{\mathbf R}^3}\d^3\bbox x
      \sum_\mu\left[(R(\bbox x) \bbox u_\mathrm{e})\cdot(\partial_\mu
      R(\bbox x) \bbox u_\mathrm{o})\right]^2\\
&&-\displaystyle
\frac{2\beta'}{a}\int_{{\mathbf R}^3}\d^3\bbox x
      \sum_\mu\left((\partial_\mu R(\bbox x) \bbox u_\mathrm{e})
      \cdot(\partial_\mu R(\bbox x) \bbox u_\mathrm{e})
     +\left(\mathrm{even}\leftrightarrow\mathrm{odd}\right)\right),
\end{array}
\end{equation}
$a$ being the lattice spacing.

If we choose, for example, the values $\bbox u_\mathrm{e}=(1,0,0)$, $\bbox
u_\mathrm{o}=(0,1,0)$, we can write 
\mbox{$\left[(R \bbox u_\mathrm{e})
\cdot(\partial_\mu R \bbox u_\mathrm{o})\right]^2=
\left[(R^T\partial_\mu R)_{12}\right]^2$}, and given that
$R^T\partial_\mu R$ is antisymmetric one can readily obtain, after
some algebra,
\begin{equation}
{\mathcal S}=\int_{{\mathbf R}^3}\d^3\bbox x
\sum_\mu\left[P(R^T\partial_\mu R)^2\right],
\label{HCONT}
\end{equation}
where $P$ is the diagonal matrix 
$(\frac{\beta}{2a}+\frac{2\beta'}{a},
               \frac{\beta}{2a}+\frac{2\beta'}{a},-\frac{\beta}{2a})$.

\end{document}